\documentclass{ws-ijmpcs}

\begin{document}

\markboth{Luigi Foschini}
{Powerful relativistic jets in spiral galaxies}

%
\catchline{}{}{}{}{}
%

\title{POWERFUL RELATIVISTIC JETS IN SPIRAL GALAXIES}

\author{LUIGI FOSCHINI}

\address{INAF Osservatorio Astronomico di Brera\\
Via E. Bianchi, 46 -- 23807 Merate (LC)
Italy\\
luigi.foschini@brera.inaf.it}

\maketitle

\begin{history}
\received{14 July 2011}
\revised{14 November 2011}
\end{history}

\begin{abstract}
The discovery of high-energy ($E>100$~MeV) $\gamma$ rays from Narrow-Line Seyfert 1 Galaxies ($\gamma$-NLS1s) has confirmed the presence of powerful relativistic jets in this class of active galactic nuclei (AGN). Although the jet emission is similar to that of blazars and radio galaxies, $\gamma$-NLS1s have some striking differences: relatively small masses ($10^{6-8}M_{\odot}$), high accretion rates ($0.1-1$ times the Eddington limit) and are generally hosted by spiral galaxies. It is now possible to study a rather unexplored range of mass and accretion rates of AGN with relativistic jets. Specifically, in this work I present some results obtained by comparing a sample of blazars and $\gamma$-NLS1s with another sample of Galactic binaries with relativistic jets (stellar mass black holes and neutron stars).
\end{abstract}

\ccode{PACS numbers: 98.54.-h, 97.80.-d}

\section{Introduction}	
During the latest decades, we have observed powerful relativistic jets from active galactic nuclei (AGN), Galactic binaries (stellar mass black holes, neutron stars and cataclysmic variables), and Gamma Ray Bursts (GRBs). Supersonic jets are also visible in protostellar systems. Despite of the diffusion of jets in almost all types of accreting astrophysical systems, it seemed that galaxies made some preference: early observations of the host galaxies of AGN with relativistic jets displayed only elliptical shapes\cite{para1,para2,para3,para4}. Only a few counterexamples (i.e. jets in spiral galaxies) were found in the past. The turning point occurred a couple of years ago with the discovery -- by means of the \emph{Fermi} satellite -- of GeV $\gamma$ rays from narrow-line Seyfert 1 galaxies ($\gamma$-NLS1)\cite{abdo1,abdo2,fosco1}. This class of AGN is composed of sources generally hosted by spiral (barred) galaxies\cite{crenshaw,deo} and, according to the common paradigm, without jets. Therefore, although some early radio observations had suggested the presence of some jet emission in a handful of cases\cite{grupe,komossa}, the discovery of GeV $\gamma$ rays was somehow a surprise. It confirmed that also this class of AGN can develop powerful relativistic jets, like blazars and radio galaxies (see Ref.~\refcite{fosco2} for a review). 

Presently, seven $\gamma$-NLS1s are known and surely one of them is hosted by a spiral galaxy\cite{zhou1,anton,fosco2}. Work is in progress to confirm the spiral hosts for the remaining $\gamma$-NLS1s. The fact that no NLS1 is known to be at high redshift\footnote{The 13th edition of the V\'eron-Cetty \& V\'eron catalogue of quasars and AGN\cite{veron} contains only NLS1s with $z\lesssim 1$. This is also in agreement with the findings that these sources are very young\cite{mathur}.} suggests that it is very unlikely to find this type of AGN hosted by ellipticals. 

The issue of the host is linked to how the mass can affect the generation and the power of the jet and was studied in the past particularly with the use of the radio loudness parameter ($RL=f_{\rm 5~GHz}/f_{\rm 440~nm}$), which in turn should indicate the dominance of the synchrotron emission from the jet over the radiation from the accretion disc (e.g. Ref.~\refcite{para3,para4}). Being the blazars and radio galaxies ($RL>>10$) confined in a mass range of $10^{8-10}M_{\odot}$, while Seyferts and LINERs ($RL<10$) occupy the range $10^{6-8}M_{\odot}$, it seemed that the jet needed of a large mass to be generated\cite{para3,para4}. 

The discovery of powerful relativistic jets in $\gamma$-NLS1 is now filling the region of small masses and high radio loudness (see Ref.~\refcite{fosco3} for more details). In addition, it opens also new interesting questions on the comparison of jets in AGN and in Galactic binaries. This is the topic of the present work and is complementary to Ref.~\refcite{fosco3}.

\section{Sample selection}
I have used the same sample of AGN in Ref.~\refcite{fosco3}, made of 9 BL Lac Objects, 30 flat-spectrum radio quasars (FSRQs) and 7 $\gamma$-NLS1s. For the present work, I have added a sample of Galactic objects (X-ray Binaries, XRB), made of three stellar mass black holes (GX~339$-$4\cite{corbel1}, H1743$-$322\cite{coriat}, V404~Cyg\cite{corbel2}) and two neutron stars (Aql~X-1\cite{miller}, 4U~1728$-$34\cite{migliari}). For XRB, the X-ray flux measured in different bands ($3-9$~keV in most of the cases), which samples the accretion radiation in Galactic objects, has been converted into the $2-10$~keV flux with {\tt webPIMMS}\footnote{{\tt http://heasarc.gsfc.nasa.gov/Tools/w3pimms.html}} by using a power-law model with photon index equal to 2. Then, the values have been corrected according to Ref.~\refcite{migliari} and scaled to the Eddington luminosity, calculated by assuming a mass of $1.4M_{\odot}$ for neutron stars and $5.8M_{\odot}$, $10M_{\odot}$, and $11.7M_{\odot}$ for GX~339$-$4\cite{gallo}, H1743$-$322\cite{coriat}, and V404~Cyg\cite{gallo}, respectively.

The radio emission has been converted into jet power by assuming the classical relationship\cite{blandford,kording} according to which $P_{\rm jet}\propto L_{\rm core,radio}^{12/17}$. For the sake of simplicity and homogeneity with the AGN sample, I have used the Eq.~(2) of Ref.~\refcite{fosco3}, taking into account that the radio emission from jets has a rather flat spectral index ($\alpha_{\rm radio}\sim 0$). The luminosities have been calculated by assuming the distances and the $N_{\rm H}$ reported in Refs.~\refcite{coriat,gallo,migliari}. 

\begin{figure}[!t]
\centerline{\includegraphics[angle=270,scale=0.4]{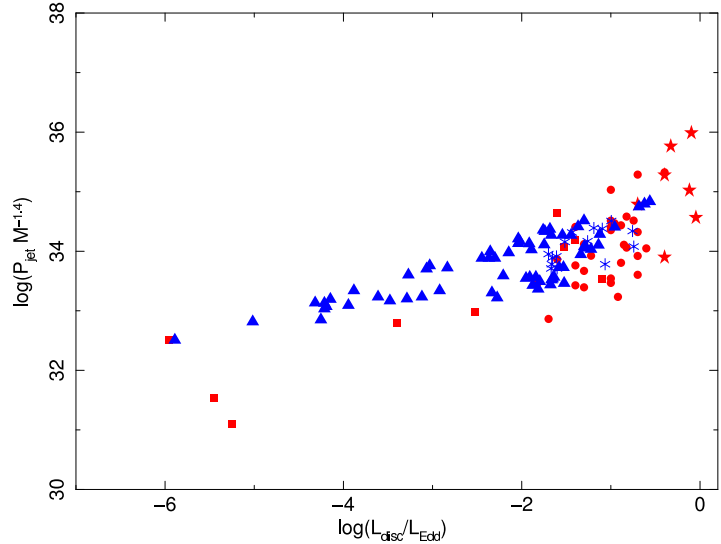}}
\vspace*{8pt}
\caption{Jet power per unit of $M^{1.4}$ vs accretion. Data from AGN are red, while XRB are blue. Squares indicate BL Lac Objects, circles are for FSRQs, and stars are $\gamma$-NLS1s. Triangles are stellar mass black holes (in different states) and asterisks are for neutron stars. See the text for more details.  \label{f1}}
\end{figure}

\section{Jet power and accretion}
As already noted in Ref.~\refcite{fosco3} (Fig.~5), the addition of $\gamma$-NLS1s to the blazars makes this sample of AGN directly comparable with a sample of jets from Galactic compact objects\footnote{In Ref.~\refcite{fosco3}, I compared Fig.~5 with an analogous figure in Ref.~\refcite{coriat} (incidentally, still Fig.~5!).}. Blazars can be compared with stellar mass black holes, while $\gamma$-NLS1s occupy the region of neutron stars. Therefore, in term of relativistic jets, $\gamma$-NLS1s are the low mass counterpart of blazars so as neutron stars are the small mass version of Galactic black holes. The difference of the jet power is about 12-13 orders of magnitude, more than the mass difference between the two samples. To remove the mass dependence, I have calculated the jet power per unit mass. Then, I have searched for correlations in the two samples by using the {\tt ASURV~v.~1.2} software package\cite{asurv1,asurv2,asurv3}. In the case of AGN, I obtain:

\begin{equation}
{\rm {\bf AGN:}} \quad \log \frac{P_{\rm jet}}{M} = (38.3 \pm 0.1) + (0.55\pm 0.06)\log \frac{L_{\rm disc}}{L_{\rm Edd}}
\label{agn}
\end{equation}

\noindent and the results for the statistical tests are: $Z=4.1$ ($P_{\rm chance}<10^{-4}$) for the Kendall's test, while the Spearman's method gives $\rho =0.6$ ($P_{\rm chance}=2\times 10^{-4}$).

In the case of Galactic binaries:

\begin{equation}
{\rm {\bf XRB:}}\quad \log \frac{P_{\rm jet}}{M} = (34.8 \pm 0.1) + (0.33\pm 0.03)\log \frac{L_{\rm disc}}{L_{\rm Edd}}
\label{xrb}
\end{equation}

\noindent with Kendall's $Z=6.7$ and Spearman's $\rho=0.7$, both with $P_{\rm chance}<10^{-4}$.

\begin{figure}[!t]
\centerline{\includegraphics[angle=270,scale=0.4]{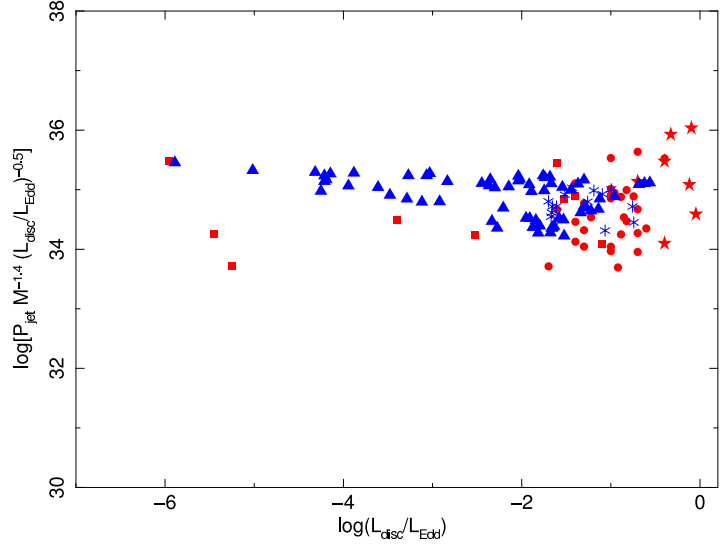}}
\vspace*{8pt}
\caption{Jet power per unit of $M^{1.4}$ and $(L_{\rm disc}/L_{\rm Edd})^{0.5}$ vs accretion. Symbols are the same of Fig.~\ref{f1}.  \label{f2}}
\end{figure}

The difference in normalization of the jet power per unit mass between AGN and XRB is now about three orders of magnitude. This difference, in excess of the difference of mass ranges, has been already noted by several authors (e. g. Refs.~\refcite{heinz,merloni}). Particularly, Ref.~\refcite{heinz} explained it by invoking a non-linear dependence on the mass, founding that $L_{\rm radio}\propto M^{17/12}$, in the hypothesis of the scale invariance of the jet. Indeed, by dividing the jet power by $M^{1.4}$, the difference in the normalization between AGN and XRB is removed (Fig.~\ref{f1}) and the correlation is now:

\begin{equation}
{\rm {\bf AGN+XRB:}}\quad \log \frac{P_{\rm jet}}{M^{1.4}} = (34.70 \pm 0.07) + (0.44\pm 0.03)\log \frac{L_{\rm disc}}{L_{\rm Edd}}
\label{all}
\end{equation}

\noindent with Kendall's $Z=9.1$ and Spearman's $\rho=0.7$, both with $P_{\rm chance}<10^{-4}$. 

Interestingly, Fig.~\ref{f1} shows also that the most efficient engines are the $\gamma$-NLS1s, with the greatest jet power per unit $M^{1.4}$ (also per unit mass): small, but nasty. This is agreement with the theory that the central engine of AGN with jets is able to distribute more or less equal power between accretion and ejection\cite{rawlings,ghise3} and, therefore, since $\gamma$-NLS1s have the greatest accretion, they are at the top end of the distribution of specific jet power. In addition, the fact that $\gamma$-NLS1s and FSRQs have very high jet powers and accretion rates, and that the Blandford-Znajek mechanism tends to saturate in a radiation-pressure dominated regime\cite{ghosh} suggest that the jet needs of an additional contribution, perhaps like Blandford-Payne or other hybrid mechanisms\cite{fosco3}. The factor 0.44 of the accretion luminosity displayed in Eq.~(\ref{all}) suggests that it is possible to remove the residual dependence shown in Fig.~\ref{f1} by dividing the specific jet power by the term $(L_{\rm disc}/L_{\rm Edd})^{0.5}$. This is indeed what is shown in Fig.~\ref{f2}. This is a fact. Its interpretation is much more cumbersome. 

This term indicates something that is not clear yet. It could point to the efficiency $\eta$ of a spherical accretion flow\cite{frank} and, thus, could emphasize the importance of Bondi accretion\cite{balmaverde,narayan2}. Another option is that it is proportional to the efficiency $\eta$ of the accretion disc, which in turn would point to the presence of some advection. Not necessarily an advection-dominated accretion flow (ADAF), but also a slim disc for the greatest rates. The hypothesis is that AGN could have composite discs, as like as XRB\cite{narayan1}, with a standard disc plus an advective part, which could be ADAF or slim depending on the accretion rate. The possibility that $(L_{\rm disc}/L_{\rm Edd})^{0.5}\propto \eta$ is appealing, because the basic concept is that high disc efficiency means high number of photons that can act both as direct and indirect (e.g. through the ionization of the broad-line region, BLR) source of seed photons to efficiently cool the relativistic electrons of the jet. It is known\cite{elitzur} that the radial scale of the BLR depends on $(L_{\rm disc}/L_{\rm Edd})^{0.5}$ and, then, the slope factor in Eq.~(\ref{all}) could be a direct indicator of the seed richness of the environment close to the central black hole (and, this time, indirectly linked to the radiative efficiency of the disc). Obviously, such an argument could be of some interest for AGN, but not for XRB that do not have a BLR. 

On the other hand, it is worth noting that the factor $0.44$ in Eq.~(\ref{all}) is just the average of the factors $0.55$ in Eq.~(\ref{agn}) and $0.33$ in Eq.~(\ref{xrb}). The physical explanation of these factors could be indeed different, indicating a separated dependence on the disc luminosity in Eddington units, although the final aim is the same. In the case of AGN, the factor $0.55$ could be linked to the BLR (and indirectly to the disc), while in the case of XRB, the factor $0.33$ could (directly) be the average for the efficiency of advective discs in different states. It is necessary to take into account that while FSRQs evolve into BL Lacs through gigayears, Galactic black holes change their states on human time scales. $\gamma$-NLS1s are on a different evolutive track as like as neutron stars are not linked to stellar mass black holes. Therefore, if the different slopes indicate different relationships (direct or indirect) with the radiative efficiency, this could be due to the different time scales.

Hypotheses, just hypotheses. How exactly the accretion disc contributes to the jet power is still to be understood. However, despite all these differences and unknowns, Fig.~\ref{f2} confirms the idea that the jets developed by all these systems are basically the same thing. 

\section*{Acknowledgments}
I would like to thank J.~M. Paredes, the LOC and the SOC of HEPRO3 for having organized a really pleasant conference and having given me time to express my ideas. Thank you also to F. Tavecchio, G. Ghisellini, and L. Maraschi for amusing and stimulating discussions and to S. Heinz for an interesting comment to my talk.

\end{document}